\documentclass[aip,jap,reprint,amsmath,amssymb,floatfix]{revtex4-2}

\usepackage{graphicx}
\usepackage{bm}
\usepackage{xcolor}
\usepackage{placeins}
\usepackage{hyperref}
\hypersetup{colorlinks=true,citecolor=blue,linkcolor=blue,urlcolor=blue}

\begin{document}

\title{A non-local power-law baseline for transient flux penetration in
YBCO films: parameter degeneracy and constitutive limitations}

\author{O. A. Hern\'andez-Flores}
\email{ohernandez.ciencias@uabjo.mx}
\affiliation{Universidad Aut\'onoma Benito Ju\'arez de Oaxaca,
Av.\ Universidad S/N, Ex-Hacienda 5 Se\~nores, 68120 Oaxaca de Ju\'arez,
Oaxaca, Mexico}

\author{C. Romero-Salazar}
\affiliation{Universidad Aut\'onoma Benito Ju\'arez de Oaxaca,
Av.\ Universidad S/N, Ex-Hacienda 5 Se\~nores, 68120 Oaxaca de Ju\'arez,
Oaxaca, Mexico}

\author{F. P\'erez-Rodr\'iguez}
\affiliation{Instituto de F\'isica, Benem\'erita Universidad Aut\'onoma de
Puebla, Edificio IF-1, Ciudad Universitaria, 72570 Puebla, Puebla, Mexico}

\date{\today}

\begin{abstract}
We assess the extent to which a non-local thin-film model with a bounded,
constant-exponent power-law constitutive relation can reproduce high-speed
magneto-optical measurements of transient flux penetration in a
$3\times3$~mm$^2$ YBa$_2$Cu$_3$O$_{7-\delta}$ film. The calculation is performed
forward in physical units and is used as a controlled baseline for judging
which observables can be represented by this constitutive family. The critical
sheet current is constrained by the final penetration geometry, whereas the
criterion field and creep exponent are constrained by the turnover and
relaxation times. The calibration reveals an intrinsic parameter degeneracy:
during the transient, the turnover time is governed primarily by the product
of the criterion field and the saturation parameter, equivalently by the
saturated sheet resistivity, so the two quantities cannot be identified
independently. Once the degeneracy is parameterized, the model captures the
observed chronology and geometry, including a guarded-current turnover near
22~ms, a relaxation constant of 41.2~ms compared with the measured 37.5 and
43.7~ms, and a late maximum settling approximately 0.92~mm from the film edge.
The amplitude does not follow. At the stated spatial resolution the transient
maximum is $1.90J_{c0}$ and the peak-to-225-ms ratio is 2.13, compared with the
reported ratio of 1.20. The excess is robust against the creep exponent,
movement along the degenerate parameter direction, changes in the Kim field
dependence, and mesh refinement. Matching the experimental ratio through the
criterion field would require a value approximately three orders of magnitude
below the conventional scale, whereas matching it by spatial exclusion would
remove the edge peak itself. We also decompose the computed electric field
into irrotational and solenoidal parts and quantify the boundary layer in which
spectral and finite-difference representations of Faraday's law disagree.
These results establish a reproducible non-local power-law baseline and show
that the remaining discrepancies are limitations of the bounded,
constant-exponent constitutive family under the stated calibration, not of
non-local thin-film electrodynamics in general.
\end{abstract}

\maketitle

\section{Introduction}\label{sec:introduction}
Magnetic-flux penetration in type-II superconductors is controlled by the
competition among the Lorentz force, vortex pinning, thermal activation, and
dissipation associated with vortex motion. Under slowly varying conditions,
critical-state models describe a wide range of long-time current and induction
profiles, including the field dependence introduced by Kim-type critical
currents \cite{Kim1962,Kim1963,McDonald1996,Shantsev1999}. These models specify
admissible magnetic states but do not uniquely determine the electric field or
the rate at which one state evolves into another. A transient calculation
therefore requires, in addition to the critical current, a constitutive
relation for vortex motion.

The distinction is especially important in thin films subjected to a
perpendicular field. Demagnetization is strong and the relation between the
sheet current and perpendicular induction is intrinsically non-local: current
at one position contributes to the magnetic field over the entire sample.
Edges, corners, and discontinuity lines consequently influence the evolution
throughout the film. Stream-function formulations combined with Fourier-space
Biot--Savart operators retain this non-local electrodynamics and permit
efficient simulations of flux creep in realistic planar geometries
\cite{Vestgarden2012,Vestgarden2013,Prigozhin2018}. A local slab model cannot
serve as a quantitative substitute for this geometry.

Time-resolved magneto-optical imaging provides an unusually stringent test of
such models. The measured perpendicular induction can be inverted to obtain
the evolving sheet-current distribution
\cite{Jooss2002,Johansen1996,Wijngaarden1996}. Wells \textit{et al.} applied
high-speed imaging to a $3\times3$~mm$^2$ YBa$_2$Cu$_3$O$_{7-\delta}$ film and
resolved a pronounced current maximum near the edge, a turnover during the
field ramp, and a subsequent stage in which the maximum moved inward and
relaxed toward a Kim-like profile \cite{Wells2016,Wells2017}. They reported a
turnover about 22~ms after penetration, relaxation constants of 37.5 and
43.7~ms for maxima originating at opposite edges, and a transient peak about
20\% above the value reached over the observation window. These measurements
constrain timing, geometry, and amplitude simultaneously and thus permit more
than a qualitative comparison.

A common macroscopic description represents creep by a power-law electric
field, $E\propto(J/J_c)^n$. The law is computationally convenient and can
approximate a restricted portion of an experimental $E$--$J$ curve. For rapid
ramps, however, an unbounded power law can generate extreme electric fields
and stiffness when the current becomes strongly overcritical. We therefore
consider a bounded form that reduces to the standard power law for $J\ll J_c$
and approaches a finite resistive branch at high current. The resulting model
contains three principal calibration quantities: the critical sheet current,
the criterion electric field, and the creep exponent, together with a
saturation parameter that controls the high-current branch.

The physical and methodological question of this work is how far this
bounded, constant-exponent constitutive family can be pushed when the
non-local thin-film geometry, experimental waveform, and dimensional scales
are all retained. In particular, we ask whether a staged calibration to the
penetration depth and timing observables can also predict the transient
amplitude. This is an overdetermined test: after three parameters have been
constrained by three observables, the remaining amplitude provides an
independent assessment of the constitutive law. The purpose is not to present
the bounded power law as a complete microscopic description, but to establish
a quantitative baseline against which more state-dependent vortex-mobility
models can be assessed.

Three related issues arise. First, the calibration may be non-identifiable if
different parameters enter the driven regime only through a common
combination. We show that the turnover time is governed primarily by the
product of the criterion field and saturation parameter, so fitting those
quantities separately is ill posed. Second, a current maximum close to a sharp
film boundary is meaningful only at a stated spatial resolution. We quantify
the layer in which spectral and finite-difference representations cease to
agree and report guarded rather than raw edge maxima. Third,
magneto-optical inversion reconstructs current from induction but does not
supply the electric field. A forward calculation gives that field directly,
allowing the Helmholtz decomposition
\begin{equation}
 \bm E=\bm E_p+\bm E_i=-\nabla\phi+\nabla\times(\psi\hat{\bm z}),
 \label{eq:helm-intro}
\end{equation}
to separate charge-related and induction-related contributions.

The contribution of the paper is therefore a controlled assessment of a
specific constitutive baseline. We identify the dissipative parameter
combination constrained by the transient, show which timing and geometric
observables the bounded power law captures, demonstrate that the amplitude
discrepancy is robust within this family, and establish the spatial-resolution
limits of the edge current and electric-field reconstruction. The conclusions
are deliberately restricted to the bounded, constant-exponent power-law
model; they do not imply that non-local thin-film electrodynamics as a whole
is unable to describe the experiment.

The remainder of the paper is organized as follows. Section~\ref{sec:formulation}
presents the thin-film equations, bounded power law, Fourier operators,
physical scales, applied waveform, and observables. Section~\ref{sec:numerics}
describes the numerical implementation and verification. The parameter
degeneracy is established in Sec.~\ref{sec:degeneracy}. Section~\ref{sec:results}
presents the staged calibration, successful predictions, amplitude
limitation, ratio decomposition, and electric-field analysis.
Section~\ref{sec:discussion} discusses the physical scope of the baseline and
the measurements needed to go beyond it. Section~\ref{sec:conclusions}
summarizes the findings.
\FloatBarrier
\section{Formulation}\label{sec:formulation}
\subsection{Thin-film problem}
A film of thickness $d$ occupies a simply connected domain $\Omega$ in the
plane $z=0$ and a uniform field $\bm H_a=H_a(t)\hat{\bm z}$ is applied
perpendicular to it. The film is square, of side $2l$, so that $l$ is its
half-width; it is the length used to non-dimensionalise the problem in
Sec.~\ref{sec:scales} and it is the only geometrical scale that enters. The sheet current is written through a stream function,
\begin{equation}
 \bm j=\nabla\times(g\hat{\bm z})=(\partial_yg,-\partial_xg),
 \label{eq:jg}
\end{equation}
so that $\nabla\cdot\bm j=0$ in $\Omega$ and $\bm j\cdot\bm n=0$ on
$\partial\Omega$; we adopt the gauge $g=0$ on and outside the sample
\cite{Vestgarden2012,Prigozhin2018}.

Because the current has been reduced to a sheet density, the material law
below relates the in-plane electric field to a current per unit width, and the
coefficient connecting them is a \emph{sheet} resistivity, measured in ohms
(per square) rather than in $\Omega\,$m. It is related to the bulk resistivity
of the film by $\rho_{\rm bulk}=\rho\,d$. We write the current--voltage
relation as a bounded power law,
\begin{equation}
 \bm E=\rho\,\bm j,\qquad
 \rho=\frac{E_c}{j_c(B_z)}\,\frac{q}{1+q/\rho_{\max}},\qquad
 q=\left(\frac{j}{j_c(B_z)}\right)^{n-1},
 \label{eq:law}
\end{equation}
with $j=\lVert\bm j\rVert$ and a Kim critical sheet current
\begin{equation}
 j_c(B_z)=\frac{j_{c0}}{1+|B_z|/B_0},\qquad j_{c0}=J_{c0}d ,
 \label{eq:kim}
\end{equation}
where $B_z=\mu_0H_z$, the film carrying no magnetisation of its own. In
Eq.~\eqref{eq:law}, $\bm E$ is in V\,m$^{-1}$ and $\bm j$ in A\,m$^{-1}$, so
$\rho$ is in $\Omega$; $E_c$ is in V\,m$^{-1}$, $j_c$ in A\,m$^{-1}$, and both
$q$ and $\rho_{\max}$ are dimensionless.

For $j\ll j_c$ the law reduces to $E=E_c(j/j_c)^n$, so $E_c$ is the criterion
field; for $j\gg j_c$ it saturates at $\rho\to\rho_{\max}E_c/j_c(B_z)$.

It is worth being explicit about what $\rho_{\max}$ is, because the symbol
invites the wrong reading. It is a pure number, the factor by which the
saturated branch exceeds $E_c/j_c$; the resistivity of that branch is
\begin{equation}
 \rho_{\rm sat}\equiv\frac{\rho_{\max}E_c}{j_{c0}} \quad [\Omega] ,
 \label{eq:rhosat}
\end{equation}
quoted at zero local induction. The ceiling inherits the field dependence of
$j_c$ and so is not constant: it equals $\rho_{\max}E_c/j_c(B_z)$ and rises
with $|B_z|$. Throughout the paper $\rho_{\max}$ is called the saturation
\emph{parameter} and $\rho_{\rm sat}$ the saturation \emph{resistivity}; only
the latter carries units. We write the ceiling with its own parameter rather
than tying it to $E_c$, as a bounded law of the form $q/(1+q)$ does, because
the two play different physical roles and, as Sec.~\ref{sec:degeneracy} shows,
conflating them hides a degeneracy rather than removing it.

The argument of Eq.~\eqref{eq:kim} is the local induction $|B_z|$.
Implementations that shift it by a global constant, for instance
$B_z-\min_D B_z$, make $j_c$ at one point depend on the field elsewhere in the
computational box; we do not use that form.

\subsection{Fourier-space operators}
For a sheet current at $z=0$, extending $g$ by zero outside $\Omega$ gives
\begin{equation}
 \widetilde H_z-\widetilde H_a=\frac{\lVert\bm k\rVert}{2}\,\widetilde g,
 \label{eq:fourier-biot}
\end{equation}
inverted as
\begin{equation}
 g=\mathcal F^{-1}\!\left[\frac{2}{\lVert\bm k\rVert}
 \left(\widetilde H_z-\widetilde H_a\right)\right]-C(t),
 \label{eq:ginv}
\end{equation}
with the multiplier set to zero at $\bm k=0$ and $C(t)$ fixed by
$\int_{\Omega_{\rm out}}g\,\mathrm d^2r=0$. Faraday's law inside the film is
\begin{equation}
 \mu_0\dot H_z=\nabla\cdot(\rho\nabla g).
 \label{eq:faraday}
\end{equation}
The exterior value of $\dot H_z$ follows from requiring $\dot g=0$ there, by
the under-relaxed iteration of Ref.~\cite{Prigozhin2018}; its convergence is
discussed in Sec.~\ref{sec:numerics}.

\subsection{Helmholtz decomposition}\label{sec:helmholtz}
Since $\bm E=\rho\bm j$ vanishes outside $\Omega$, it is compactly supported
in the computational box and Eq.~\eqref{eq:helm-intro} can be evaluated
spectrally without a boundary artefact:
\begin{equation}
 \widetilde\phi=\frac{i\,\bm k\cdot\widetilde{\bm E}}{\lVert\bm k\rVert^2},
 \qquad
 \widetilde\psi=\frac{i\left(k_x\widetilde E_y-k_y\widetilde E_x\right)}
 {\lVert\bm k\rVert^2},
 \label{eq:helm-fourier}
\end{equation}
with both multipliers zero at $\bm k=0$, and
$\bm E_p=-\nabla\phi$, $\bm E_i=(\partial_y\psi,-\partial_x\psi)$.

Two properties make the split worth computing. Since
$(\nabla\times\bm E)_z=-\mu_0\dot H_z$, the solenoidal potential obeys
$\nabla^2\psi=\mu_0\dot H_z$: the entire dependence of $\bm E$ on the applied
ramp is carried by $\psi$, and $\bm E_p$ is by construction blind to $\dot B$.
And since $\nabla\cdot\bm E_i=0$, all of the divergence resides in $\bm E_p$,
which therefore images the transient charge whose role near discontinuity
lines was emphasised in Refs.~\cite{Brandt1995,RomeroSalazar2010}.

\subsection{Physical scales}\label{sec:scales}
Lengths are measured in units of the film half-width $l$, currents in units of
the zero-field critical sheet current $j_{c0}=J_{c0}d$. With
$\bar x=x/l$, $\bar{\bm j}=\bm j/j_{c0}$, $\bar H=H/j_{c0}$,
$\bar g=g/(l j_{c0})$ and $\bar t=t/t_0$, Eq.~\eqref{eq:faraday} becomes
$\partial_{\bar t}\bar H_z=\bar\nabla\cdot(\bar\rho\bar\nabla\bar g)$ provided
\begin{equation}
 t_0=\frac{\mu_0\,j_{c0}\,l}{E_c},
 \qquad \bar H_{\max}=\frac{B_{\max}}{\mu_0 j_{c0}},
 \qquad \bar B_0=\frac{B_0}{\mu_0 j_{c0}} .
 \label{eq:t0}
\end{equation}
Note that $\bar H_{\max}$ and $t_0$ are controlled by different combinations:
the first fixes how deeply the flux penetrates and depends only on $J_{c0}d$;
the second sets the clock and is the only place where $E_c$ enters. This
separation is what makes a staged calibration possible.

\subsection{Applied waveform}
The magnet of Refs.~\cite{Wells2016,Wells2017} ramps at
$\dot B_a=(8.1\pm2.3)$~T\,s$^{-1}$ and settles at $0.1$~T after about
$30$~ms, slowing near the end to avoid overshoot. We represent this by
\begin{equation}
 B_a(t)=B_{\max}\left[1-\cos(\omega t)\,e^{-t/\tau_r}\right],
 \qquad \tau_r=\frac{B_{\max}}{\dot B_a(0)}=12.35\ \mathrm{ms},
 \label{eq:protocol}
\end{equation}
with $\omega\tau_r=1.1\pi/7$, which reproduces the initial rate exactly and
reaches $97\%$ of $B_{\max}$ at $30$~ms. The subscript distinguishes this
constant of the applied ramp from the relaxation constant $\tau$ of the
film's own response, defined next; they differ by a factor of three and
confusing them would make nonsense of the calibration.

\FloatBarrier

\subsection{The two timing observables}\label{sec:observables}
Two times are extracted from each calculation and both are compared with the
measurements; since the whole calibration rests on them, we define them
precisely.

The first is the \emph{turnover time} $t_{\rm cross}$. Let
\begin{equation}
 J_{\max}(t)=\max_{\bm r\in\Omega_g}\lVert\bm J(\bm r,t)\rVert
 \label{eq:Jmax}
\end{equation}
be the largest current density in the film at time $t$, the maximum being
taken over the guarded region $\Omega_g$ defined in
Sec.~\ref{sec:verification} rather than over the whole film. During the ramp
$J_{\max}$ grows; once the applied field stabilises it decays. We define
\begin{equation}
 \left.\frac{\mathrm dJ_{\max}}{\mathrm dt}\right|_{t=t_{\rm cross}}=0,
 \qquad
 J_{\max}(t_{\rm cross})=\max_t J_{\max}(t) ,
 \label{eq:tcross}
\end{equation}
that is, $t_{\rm cross}$ is the instant at which the curve
$J_{\max}(t)$ attains its largest value and growth turns into decay. It is the model's
counterpart of the transition Wells \textit{et al.} identify from
discontinuities in the peak current and the flux-front velocity, and which
they place about $22$~ms after penetration \cite{Wells2017}. In practice $t_{\rm cross}$ is read off the sampled curve $J_{\max}(t)$ as the
location of its largest sample, so its precision is that of the output-time
grid; this is the origin of the small steps visible in Fig.~\ref{fig:map}(b).
The quantity is reported only when $J_{\max}$ genuinely turns over. If the
current rises and then flattens without a maximum --- which happens when the
overshoot is suppressed --- the largest sample falls at an arbitrary point of
the plateau and carries no meaning, and no value is quoted.

The second is the \emph{relaxation constant} $\tau$. After the turnover the
position $p(t)$ of the current maximum along the central line moves inward and
settles, and the motion is fitted, as in the experiment, by
\begin{equation}
 p(t)=p_\infty-A\,e^{-(t-t_{\rm cross})/\tau} ,
 \label{eq:taufit}
\end{equation}
with $p$ measured from the film edge.

Which maximum is followed has to be said explicitly, because the profile
carries more than one. Besides the migrating peak there is a maximum that
remains within a few tens of microns of the edge and does not move, and at
late times the plateau between them develops further weak maxima; up to five
local maxima were counted outside the guard in the calibrated run. Wells
\textit{et al.} distinguish the two principal features, describing "a sloped
plateau between this peak and a shoulder, whose onset remains close to the
initial peak position at the sample edge", and fit the peak. We therefore
track the \emph{innermost} significant maximum, not the largest, and locate it
to sub-cell precision by a parabola through its three neighbouring points.
Both refinements are necessary rather than cosmetic. Following the largest
maximum instead makes the tracked position jump discontinuously, by almost a
millimetre, at the instant the migrating peak overtakes the stationary one,
and an exponential fitted across that step returns the timing of the step
rather than a relaxation constant. Reporting the position at grid resolution
makes $p(t)$ a staircase of some twenty-five steps over its whole travel, from
which the fitted $\tau$ scatters by tens of per cent and ceases to be
monotonic in $n$. That the resulting trajectory runs parallel to the flux
front, some $0.1$~mm behind it throughout --- as Wells \textit{et al.} also
report --- is an independent check that it is the intended feature that is
being followed.

The three fit parameters are the
relaxation constant $\tau$; the limiting position $p_\infty$, that is the
distance from the edge at which the maximum finally settles; and the amplitude
$A=p_\infty-p(t_{\rm cross})$, the total distance it travels inward from the
turnover onwards. Of these, $\tau$ and $p_\infty$ are compared with the
measurements below. The peak is tracked on the left half of the central line
only: the profile carries one maximum near each edge, and following the global
maximum would let a negligible left--right asymmetry decide which of the two is
reported.

Wells \textit{et al.} report $\tau=37.5$ and $43.7$~ms for the peaks
originating at opposite edges of their film; the spread between the two is what
sets the tolerance on our calibration of the creep exponent. They place the
final position at about $0.9$ to $1.1$~mm from the edge.

The ratio $\tau/t_{\rm cross}$ is used repeatedly below because it is
dimensionless and both times scale with $t_0$, so it is independent of $E_c$
and of the overall clock of the problem.

\FloatBarrier
\section{Numerical method and its limits}\label{sec:numerics}
\subsection{Discretisation}
The film $|\bar x|,|\bar y|\le1$ is embedded in a box $[-1.25,1.25]^2$
discretised on a periodic $N\times N$ grid with the sample boundary on grid
nodes (Appendix~\ref{app:grid}); $N=144$, $200$, $288$ and $400$ were used.
Spatial derivatives in Eq.~\eqref{eq:faraday} use fourth-order finite
differences and the non-local operators use two-dimensional FFTs, filtered
with $\mathcal G(\bm k)=\exp(-\tfrac12|\bm k|^2\sigma^2)$. The same difference
operator is used during integration and in post-processing, so the reported
sheet current is the one the constitutive law sees.

Two choices differ from common practice and both matter for reproducibility.
The filter width $\sigma$ is fixed in physical units, $\sigma=0.03\,l=45$~$\mu$m,
rather than tied to the mesh; and the exterior iteration is restarted from a
fixed guess at every evaluation of the right-hand side and run to a
tolerance, so that the right-hand side is a function of $(\bar t,\bar g)$
alone, as adaptive integrators assume. A fixed number of sweeps from a warm
start makes the computed derivative depend on the history of solver calls,
including rejected steps. Time integration uses \texttt{ode23} with Anderson
mixing on the exterior iteration; the converged solution is reproducible to
all printed digits across repeated runs.

\subsection{Convergence of the exterior iteration}\label{sec:exterior}
The iteration does not converge to an arbitrary tolerance, and it is worth
saying why rather than quietly choosing a fixed number of sweeps. Its residual
--- defined as $R_{\rm out}$ in Eq.~\eqref{eq:resout} below --- falls by two
orders of magnitude over the first few sweeps and then stalls, because the
correction carries the factor $\lVert\bm k\rVert/2$, which vanishes as
$\bm k\to0$: the largest-scale exterior mode is barely corrected at all.
Measured contraction factors are $0.81$ over sweeps $3$--$9$ but $0.996$
asymptotically, so pushing $R_{\rm out}$ from $2\times10^{-2}$ to $10^{-3}$
would take some $750$ sweeps. We therefore stop on stagnation and report the
value reached.

Whether the residual left over matters was tested directly rather than
assumed: between $6$ and $60$ sweeps it changes by a factor of five while the
turnover time is unchanged and the amplitude and peak electric field vary by
$1.4\%$ and $2.1\%$. The stray exterior current left by a short iteration does
not affect any conclusion.

This also explains, without appeal to convention, why implementations in the
literature use a fixed small number of exterior corrections.

\subsection{Verification}\label{sec:verification}
Three residuals are monitored at every output time. They test different parts
of the calculation and it is worth keeping them apart.

\emph{(i) Completeness of the Helmholtz decomposition.}
\begin{equation}
 R_H=\frac{\lVert\bm E_p+\bm E_i-\bm E\rVert}{\lVert\bm E\rVert}
 \label{eq:resH}
\end{equation}
tests whether the two spectral inversions of Eq.~\eqref{eq:helm-fourier}
return, when added back together, the field they were computed from. It probes
the decomposition alone and nothing else, and it reaches machine precision,
$R_H\sim10^{-16}$. Trivial as the test looks, it is the one that matters most
in practice: a sign error in the multiplier for $\psi$ leaves $|\bm E_i|$
correct while reversing its direction, so every map looks plausible and only
$R_H$ betrays the mistake, rising to $\approx2$.

\emph{(ii) Convergence of the exterior iteration.}
\begin{equation}
 R_{\rm out}=\frac{\langle|\dot{\bar g}|\rangle_{\Omega_{\rm out}}}
                  {\langle|\dot{\bar g}|\rangle_{D}}
 \label{eq:resout}
\end{equation}
measures how well the iteration achieves what it exists for, namely
$\dot g=0$ outside the film, the averages being taken over the exterior region
and over the whole computational box respectively. For the reasons given in
Sec.~\ref{sec:exterior} it stalls rather than converging: it averages
$3.7\times10^{-2}$ over the interval, with a maximum of $8.5\times10^{-2}$ at
the very end, where the drive has died away and $\dot g$ is negligible
everywhere so that the ratio is taken between two small numbers. The
observables are insensitive to it.

\emph{(iii) Consistency of the two discrete operators.} Inside the film
Faraday's law gives $-(\nabla\times\bm E)_z=\mu_0\dot H_z$ identically, since
$\bm E=\rho\,\nabla\times(g\hat{\bm z})$ by construction. Evaluating the left
side spectrally and the right side with the finite-difference operator used in
the time integration therefore does not test the physics; it tests whether the
two discretisations agree on the fields actually produced. With both sides
unfiltered and normalised by the full-film norm of $\dot H_z$,
\begin{equation}
 R_{FD}=\frac{\lVert(\nabla\times\bm E)_z+\mu_0\dot H_z\rVert_S}
             {\lVert\mu_0\dot H_z\rVert_\Omega} ,
 \label{eq:resFD}
\end{equation}
where $S$ is the region over which the norm is taken. Its value depends
strongly on that choice: $R_{FD}=0.12$--$0.18$ when $S$ is the guarded
interior, but $2.2$ when $S$ is the whole film.

That contrast is the central numerical result of this section. Within a few
cells of the film boundary the two discrete representations of the same
quantity differ by more than the signal: the spectral operator rings against
the jump in $\bm j$ and the finite-difference stencils become one-sided across
it. Neither is reliable there. Consequently the raw maximum of $|\bm J|$,
which falls inside that layer, is not a numerically defined quantity, and all
amplitudes below are reported with a stated guard of $0.09\,l=135$~$\mu$m
around the boundary. With $\sigma$ and the guard both fixed in physical units,
the guarded amplitude varies by $3\%$ and non-monotonically over
$N=144$ to $400$; with $\sigma$ and the guard tied to the mesh, as is usual, it
drifts upward without converging, because refining the grid narrows the
excluded layer rather than resolving the peak.

\subsection{A caution about relative residuals}
Every diagnostic above is a ratio, and in this problem the natural
denominators vanish in the regimes of interest: there is no current at
$t=0$, no dynamics once the drive has decayed, and negligible dissipation deep
inside the film where $j\ll j_c$. A residual normalised by a regional norm
therefore reports division by noise and raises false alarms precisely where
the solution is well behaved. We normalise by global scales with an absolute
floor throughout. The point is practical rather than deep, but it cost us
considerable time and we expect it would cost the same to anyone
re-implementing the method.

\FloatBarrier
\section{Degeneracy of the dissipation parameters}\label{sec:degeneracy}
Two of the parameters of Eq.~\eqref{eq:law}, the criterion field $E_c$ and the
dimensionless saturation parameter $\rho_{\max}$, cannot be determined
independently from
the transient.

\begin{figure*}[!tb]
\centering
\includegraphics[width=\textwidth]{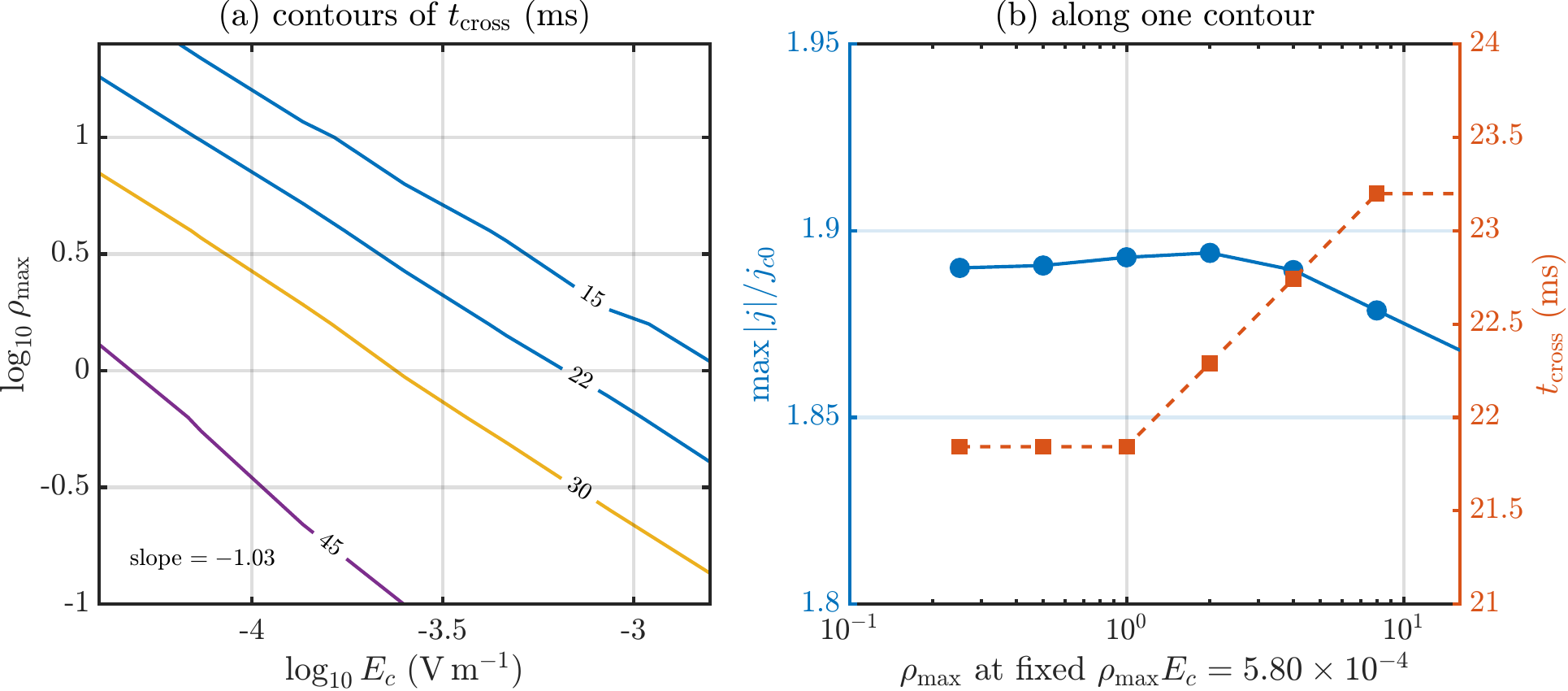}
\caption{Degeneracy of the two dissipation parameters.
(a) Contours of the turnover time $t_{\rm cross}$ in the
$(\log_{10}E_c,\log_{10}\rho_{\max})$ plane at fixed $n=64$, labelled in
milliseconds. They are straight and parallel, the $22$~ms contour having slope
$-1.032$ over $1.6$ decades, so $t_{\rm cross}$ is a function of the product
$\rho_{\max}E_c$ and the two parameters cannot be determined separately from
the timing.
(b) What happens on moving along one such contour, $\rho_{\max}E_c$ held at
$5.80\times10^{-4}$, as $\rho_{\max}$ runs over a factor of $64$: the peak
current (left axis) varies by $1.4\%$, while $t_{\rm cross}$ (right axis)
drifts by $6\%$, so the invariance is close but not exact. Note the suppressed
zero on both axes of (b). The steps in $t_{\rm cross}$ are the spacing of the
output-time grid rather than structure in the solution.}
\label{fig:map}
\end{figure*}

Figure~\ref{fig:map} shows the contour of the turnover time
$t_{\rm cross}$ of Eq.~\eqref{eq:tcross}, at the measured value of $22$~ms, in
the
$(\log E_c,\log\rho_{\max})$ plane. Over $1.6$ decades it is a straight line of
slope $-1.03$, that is
\begin{equation}
 t_{\rm cross}\simeq t_{\rm cross}(\rho_{\max}E_c)
 = t_{\rm cross}(j_{c0}\,\rho_{\rm sat}) .
 \label{eq:degeneracy}
\end{equation}
The degenerate combination is not an accident of parametrisation: since
$j_{c0}$ is fixed independently by the penetration depth, what
Eq.~\eqref{eq:degeneracy} says is that the turnover time is controlled by the
saturated resistivity alone.
The invariance is close but not exact: moving along the line
$\rho_{\max}E_c=\mathrm{const}$ over a factor of $64$ in $\rho_{\max}$ shifts
$t_{\rm cross}$ from $21.8$ to $23.2$~ms, a drift of $6\%$.
The reason is visible in the solution: at the current peak the film is
entirely on the saturated branch of Eq.~\eqref{eq:law}, where the resistivity
is $\rho_{\max}E_c/j_c$; the ratio of the peak electric field to $E_c$ equals
the current amplification to three digits. The two parameters enter that
regime only through their product.

An alternating fit that adjusts $\rho_{\max}$ against one observable and $E_c$
against another therefore does not converge: it slides along the degenerate
direction. In our first attempt it ran away to $E_c=3.0\times10^{-5}$~V\,m$^{-1}$,
where $t_0$ exceeds a second and the applied waveform is effectively a step,
while reporting that both targets had been met --- they had been met at
different points of the iteration, never simultaneously.

The creep exponent $n$ breaks the degeneracy, because it sets the shape of the
crossover between the creep and saturated branches and therefore changes the
relaxation time without changing the saturated resistivity $\rho_{\rm sat}$. Its action is
remarkably clean: over $n=18$ to $140$ at fixed $E_c$ and $\rho_{\max}$, the
turnover time is constant at $22.3$~ms to four digits and the peak amplitude
varies by $0.3\%$, while $\tau$ falls monotonically from $93$ to $29$~ms. Each of the
three parameters therefore controls essentially one observable. Because the $t_{\rm cross}$ contour is vertical in the $(\log E_c,n)$ plane,
$E_c$ is fixed by the timing almost independently of $n$ and the calibration
reduces to a one-dimensional scan along that constraint

\begin{figure}[!htb]
\centering
\includegraphics[width=\columnwidth]{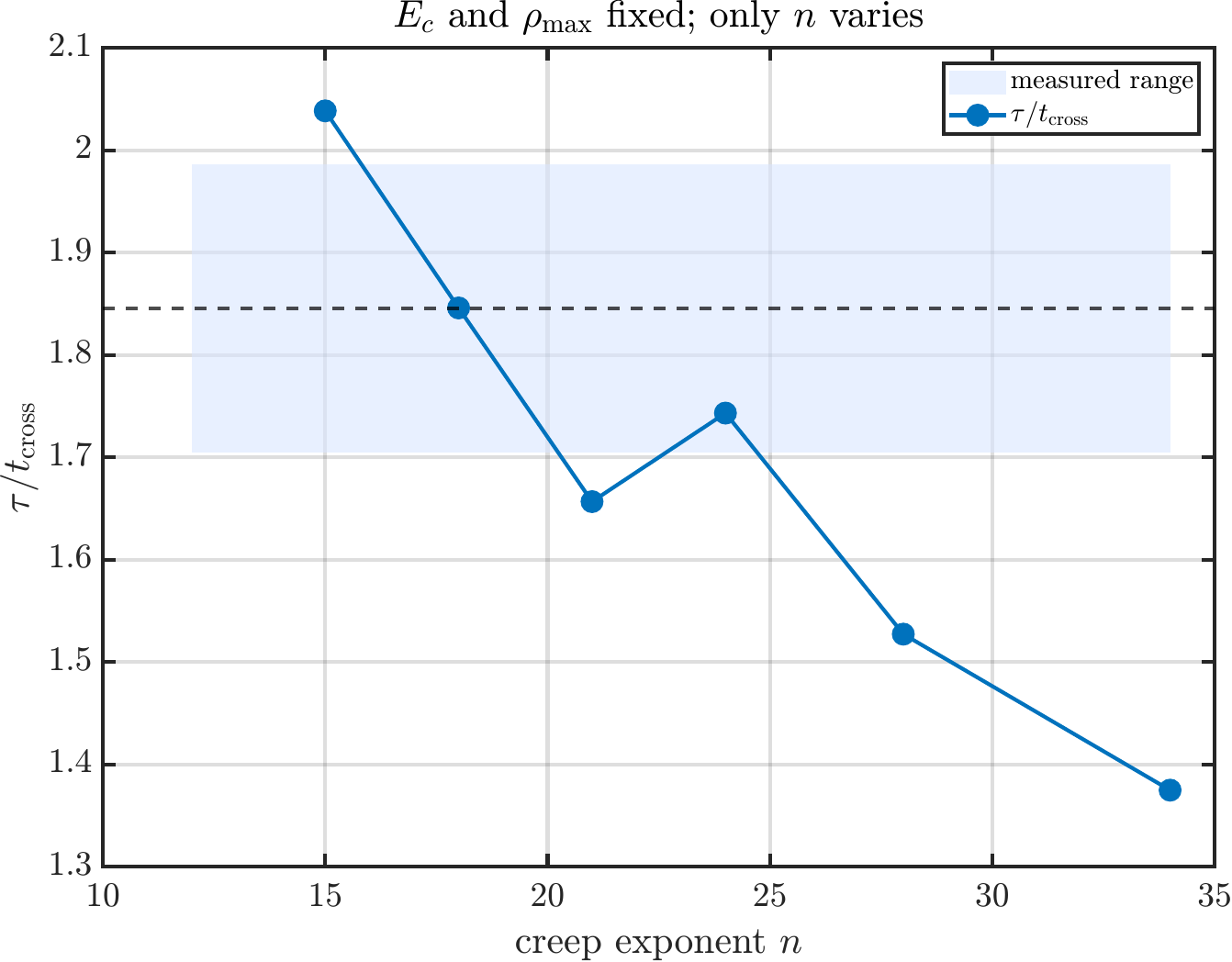}
\caption{Fixing the creep exponent. The dimensionless ratio
$\tau/t_{\rm cross}$ against $n$, with $E_c$ and $\rho_{\max}$ held at their
calibrated values and on the same $288\times288$ mesh as the reference run.
The shaded band is the range implied by the two relaxation constants measured
on opposite edges of the film, $37.5$ and $43.7$~ms, against a turnover at
$22$~ms; the dashed line is its midpoint. The curve is smooth and monotonic,
$\tau$ falling from $93$ to $29$~ms as $n$ goes from $18$ to $120$, and it
crosses the band between $n\approx56$ and $n\approx72$. Neither
$t_{\rm cross}$ nor the peak amplitude varies over this range: they are
constant at $22.3$~ms and $1.90\,J_{c0}$ to four and three digits
respectively, so the exponent is fixed by $\tau$ alone and does not disturb
the other two observables.}
\label{fig:nscan}
\end{figure}

(Fig.~\ref{fig:nscan}). It gives
\begin{equation}
 E_c=2.9\times10^{-4}\ \mathrm{V\,m^{-1}},\qquad n=64 ,
\end{equation}
The two measured relaxation constants bracket $n$ between about $56$ and
$72$; the calculation itself is smooth and monotonic in $n$ over this range,
so the uncertainty is set by the experiment and not by the model.

We note that $E_c$ so obtained is within a factor of three of the conventional
criterion field of $10^{-4}$~V\,m$^{-1}$, and $n=64$ is an ordinary exponent
for a well-pinned YBCO film at low temperature. The implied saturated resistivity, however, is
$\rho_{\rm sat}=4.1\times10^{-9}$~$\Omega$ as a sheet value, or
$\rho_{\rm sat}d=2.1\times10^{-15}$~$\Omega$\,m in bulk terms, both at zero
local induction --- some five orders of magnitude below the Bardeen--Stephen
estimate $\rho_nB/B_{c2}$. We read this
as a statement that the dissipation sampled by the transient is that of
thermally activated creep rather than free flux flow, and not as a measurement
of the flux-flow resistivity; the saturated branch of Eq.~\eqref{eq:law} is
reached at $j\approx2j_c$, far from the regime where Bardeen--Stephen applies.

\FloatBarrier
\section{Calibration and results}\label{sec:results}
\subsection{The constrained baseline parameters}
The parameter selection is staged so that each principal quantity is
constrained by the observable to which it is most sensitive, in an order that
avoids circularity. The resulting values define an experiment-informed
power-law baseline rather than a unique microscopic identification.

\emph{The critical sheet current $j_{c0}$ from the penetration depth.} For a
thin strip the flux front reaches $b=l/\cosh(\pi H_a/j_c)$, measured from the
centre \cite{BrandtIndenbom1993}. Wells \textit{et al.} show the front settling
about $1.0$~mm from the edge of their $3\times3$~mm$^2$ film, that is
$l=1.5$~mm and $b/l=1/3$, whence $\pi H_a/j_c=\operatorname{arccosh}3$, giving
$\bar H_{\max}=0.56$ and
\begin{equation}
 j_{c0}=J_{c0}d=1.4\times10^5\ \mathrm{A\,m^{-1}} .
 \label{eq:jc0}
\end{equation}
The estimate uses the Bean form, but the value is confirmed by the full
nonlinear calculation: run with Eq.~\eqref{eq:jc0} and the Kim law, the model
places the final current maximum $1.0$~mm from the edge
(Sec.~\ref{sec:right}), inside the measured range.

It is the sheet current, not the bulk current density, that this argument
constrains, and it is worth being explicit that the distinction has no
consequences for anything else in the paper. The thickness enters the
formulation only through the product $J_{c0}d$, so every dimensionless
quantity, and therefore every time and every current ratio reported below, is
unchanged by the choice of $d$. Only two derived numbers depend on it: the
bulk critical current density $J_{c0}=j_{c0}/d$ and the bulk equivalent
$\rho_{\rm sat}d$ of the saturated resistivity. The thickness of the film of
Refs.~\cite{Wells2016,Wells2017} is not stated there; the deposition work it
cites reports films between $400$ and $1000$~nm, over which range
$J_{c0}=1.4$ to $3.5\times10^{11}$~A\,m$^{-2}$ and
$\rho_{\rm sat}d=1.7$ to $4.1\times10^{-15}$~$\Omega$\,m. We quote $d=500$~nm
where a bulk figure is needed and carry the range as an uncertainty on those
two quantities alone.

This does raise a question about the sample that we can pose but not settle.
A critical current density of $10^{12}$~A\,m$^{-2}$ is quoted for a similar
film at $10$~K in Ref.~\cite{Wells2016}; combined with the thickness range
above it would give $j_{c0}$ between $4\times10^5$ and $10^6$~A\,m$^{-1}$,
three to seven times the value Eq.~\eqref{eq:jc0} requires. At such a sheet
current $\bar H_{\max}$ would fall to $0.08$--$0.2$ and the flux would barely
enter the film, which is not what the imaging shows. Either the film is
thinner than the deposition range suggests, or its critical current density at
$7$~K is well below the quoted figure, or the applied induction at the sample
differs from the nominal $0.1$~T. We proceed with the value the penetration
depth demands, since it is the one that reproduces the observed geometry, and
note the discrepancy as a property of the published data rather than of the
model.

\emph{$E_c$ and $n$ from the two times}, as in Sec.~\ref{sec:degeneracy}.

\emph{The Kim constant $B_0$ is bounded, not assumed.} It is not measured for
this film, and we adopt $B_0=0.5$~T, giving $\bar B_0=2.84$. The value is not
free, however: the observation that the current maximum detaches from the edge
and migrates inward constrains it from above. For $B_0$ beyond about $0.8$~T
the computed maximum never leaves the edge at any $j_{c0}$, so the migration
reported in Fig.~4a of Ref.~\cite{Wells2017} cannot be reproduced at all. The
reason is that the interior maximum is itself a Kim feature: the local
induction is largest at the film edge, so $j_c$ is smallest there and largest
behind the flux front, and a weak field dependence flattens the profile into a
Bean-like plateau with no interior maximum to move. The observed migration is
thus evidence for a field-dependent critical current, and it bounds $B_0$ from
above.

Within the allowed range the results are insensitive to the choice. Repeating
the calibration at $B_0=0.25$, $0.40$ and $0.50$~T, re-fixing $j_{c0}$ each
time so that the final peak position again matches $1.0$~mm, gives
$j_{c0}=1.55$, $1.43$ and $1.39\times10^{5}$~A\,m$^{-1}$; a peak amplitude of
$1.72$, $1.83$ and $1.87\,J_{c0}$; and a peak-to-asymptote ratio of $2.26$,
$2.33$ and $2.35$. The quantity on which the conclusions rest, the ratio,
moves by $3.9\%$ over that range, and $\tau/t_{\rm cross}$ by $5.6\%$. None of
the discrepancy reported in Sec.~\ref{sec:amplitude} can be attributed to the
assumed field dependence.

All parameters are collected in Table~\ref{tab:params}. Every number quoted
below follows from them with nothing further adjusted.

\begin{table}[t]
\centering
\caption{Parameters of the calibrated model. The first three are fitted, each
to the observable named; the rest are fixed by the sample or the apparatus.}
\label{tab:params}
\begin{tabular}{llll}
\hline
Quantity & Symbol & Value & Fixed by \\
\hline
Critical sheet current & $j_{c0}$ & $1.4\times10^5$ A\,m$^{-1}$ & front position \\
Dissipation field & $E_c$ & $2.9\times10^{-4}$ V\,m$^{-1}$ & $t_{\rm cross}$ \\
Creep exponent & $n$ & $64$ ($56$--$72$) & $\tau/t_{\rm cross}$ \\
Saturation parameter & $\rho_{\max}$ & $2$ (dimensionless) & degenerate with $E_c$ \\
Saturated resistivity & $\rho_{\rm sat}$ & $4.1\times10^{-9}$ $\Omega$ & Eq.~\eqref{eq:rhosat} \\
Half-width & $l$ & $1.5$ mm & sample \\
Thickness & $d$ & $500$ nm ($400$--$1000$) & assumed; see text \\
Kim constant & $B_0$ & $0.5$ T ($\lesssim0.8$) & bounded by the peak motion \\
Final induction & $B_{\max}$ & $0.1$ T & apparatus \\
Ramp rate & $\dot B_a(0)$ & $8.1$ T\,s$^{-1}$ & apparatus \\
Time scale & $t_0$ & $0.91$ s & Eq.~\eqref{eq:t0} \\
Filter width & $\sigma$ & $45$ $\mu$m & stated resolution \\
Edge guard & --- & $135$ $\mu$m & Sec.~\ref{sec:verification} \\
\hline
\end{tabular}
\end{table}

\subsection{Observables captured by the baseline}\label{sec:right}

\begin{figure}[!htb]
\centering
\includegraphics[width=\columnwidth]{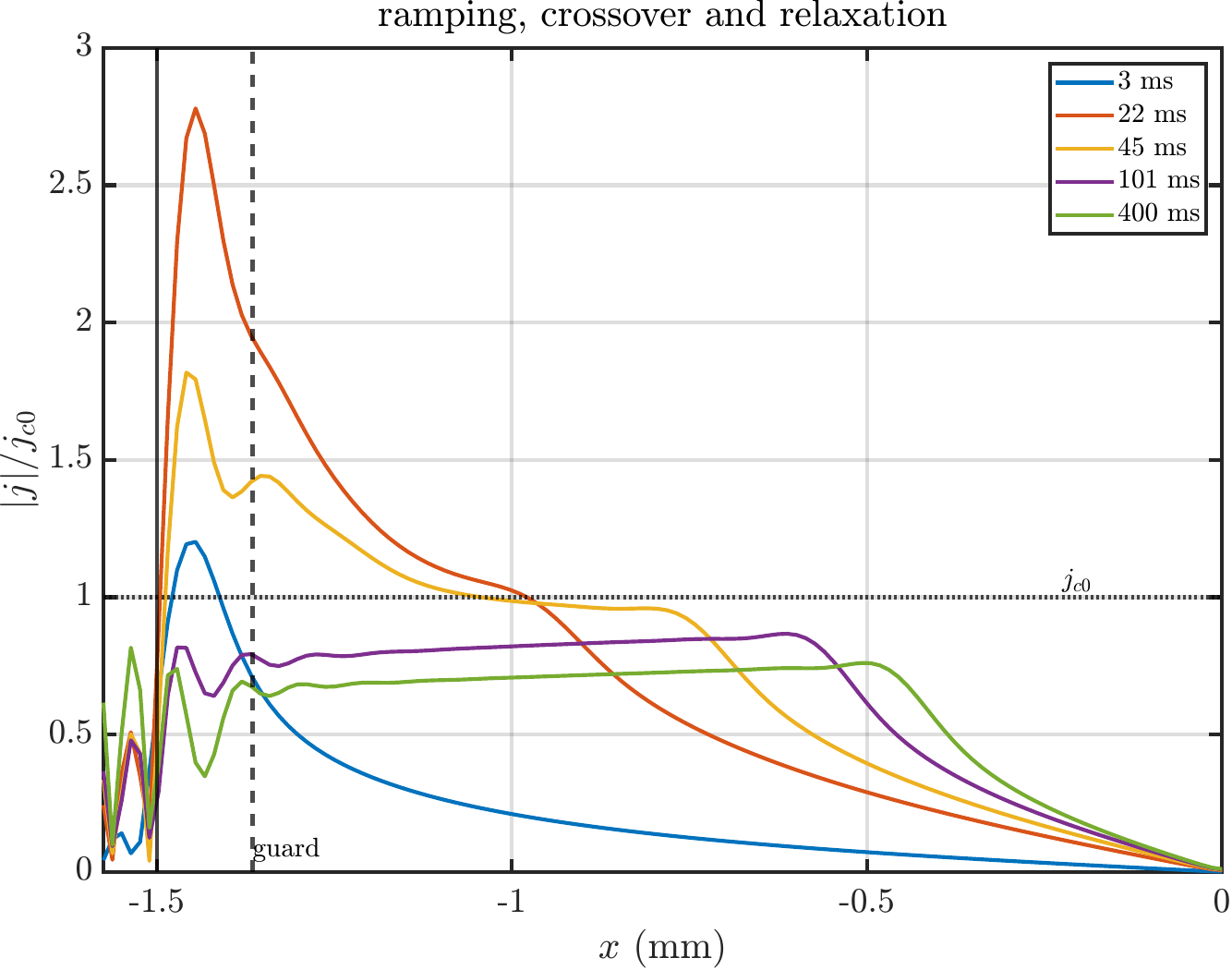}
\caption{Sheet-current profiles along the left half of the central line of the
film, at five instants of the calibrated run. The solid vertical line is the
film edge and the dashed vertical line the edge guard of $135$~$\mu$m; the
dotted horizontal line marks $j_{c0}$. Currents are normalised by the
critical sheet current; the ratio is identical to $|\bm J|/J_{c0}$, since the
thickness cancels. The three stages are visible in
sequence: screening decaying inward at $3$~ms, a large peak at the edge with a
long tail at $22$~ms, the peak detached with a shoulder behind it at $45$~ms,
and the peak settled about $1$~mm from the edge with a plateau behind it at
$101$ and $400$~ms. The oscillations beyond $-1.5$~mm lie \emph{outside} the
film, where no current should flow; they are the edge ringing quantified in
Sec.~\ref{sec:verification}, and they are the reason the peak is reported with
a guard rather than as a raw maximum.}
\label{fig:profiles}
\end{figure}

Figure~\ref{fig:profiles} shows the sheet-current profile along the central
line. The three stages reported from the imaging are recovered: a maximum
forms near the edge and grows while the field is driven; its growth stops and
a shoulder develops; the maximum then moves inward and decays while the
shoulder broadens, leaving a profile with a peak behind the flux front and a
plateau toward the edge.

Quantitatively, over $N=144$ to $400$ at fixed $\sigma$ and guard,
\begin{align}
 t_{\rm cross} &= 22.2\pm0.5\ \mathrm{ms} & &(\text{fitted; target } 22), \\
 \tau &= 41.2\ \mathrm{ms} & &(\text{measured } 37.5,\ 43.7), \\
 p_\infty &= 0.92\ \mathrm{mm} & &(\text{measured } 0.9\text{--}1.1).
\end{align}
The relaxation constant and the final peak position are consistency checks
rather than predictions: $\tau$ enters through the ratio used to fix $n$, and
$p_\infty$ through the front position used to fix $J_{c0}d$. What they show is
that the staged calibration is consistent --- fixing $J_{c0}d$ from the
Brandt--Indenbom estimate reproduces the measured final position of the peak,
and fixing $n$ from a dimensionless ratio reproduces a dimensional time within
the spread of the two measured values.

\subsection{Constitutive limitation revealed by the amplitude}\label{sec:amplitude}
The amplitude does not follow. The transient maximum reaches
\begin{equation}
 \frac{\max|\bm J|}{J_{c0}}=1.90
 \label{eq:amp}
\end{equation}
at the stated resolution, and this number is remarkably hard to move. It
varies by $0.3\%$ over $n=12$ to $34$; by $1.4\%$ along the degenerate
direction as $\rho_{\max}$ runs from $0.25$ to $16$ with $\rho_{\max}E_c$
held fixed; and by $3\%$ over the mesh sequence $N=144$ to $400$. In other
words the peak current is not merely a value we happened to obtain with one
parameter set: within this constitutive family it is essentially not
adjustable, because the parameters that could move it are pinned by the
timing observables.

The comparison with experiment must be made through the peak-to-asymptote
ratio, since Wells \textit{et al.} infer their critical current from the
long-time data rather than measuring it independently. That ratio is
\begin{equation}
 \left.\frac{J_{\rm peak}}{J(225\,\mathrm{ms})}\right|_{\rm model}=2.13
 \qquad\text{against}\qquad 1.20\ \text{measured} .
\end{equation}

Unlike the peak itself, this ratio is \emph{not} invariant along the
degenerate direction: it falls with $\rho_{\max}$ as that parameter goes
from $0.25$ to $16$. The reason is that the denominator moves while the
numerator does not. At long times $j\ll j_c$, the resistivity of
Eq.~\eqref{eq:law} reduces to $\rho\simeq E_c q/j_c$ irrespective of
$\rho_{\max}$, so the creep rate is set by $E_c$ alone; raising
$\rho_{\max}$ at fixed product lowers $E_c$, slows the creep, and lifts the
value the current has relaxed to by $225$~ms.

The dependence is accurately log-linear, with a slope of $-0.32$ per decade
of $\rho_{\max}$. Extrapolating it, the measured ratio of $1.20$ would be
reached at $\rho_{\max}\approx10^4$ and
\begin{equation}
 E_c\approx6\times10^{-8}\ \mathrm{V\,m^{-1}},
\end{equation}
three orders of magnitude below the conventional criterion field and with
$t_{\rm cross}$ already drifting away from its target. We therefore regard the discrepancy as a robust limitation of the bounded,
constant-exponent power-law family under the stated constraints, rather than
as a nearby region of its parameter space that was not searched.

\begin{figure}[!htb]
\centering
\includegraphics[width=\columnwidth]{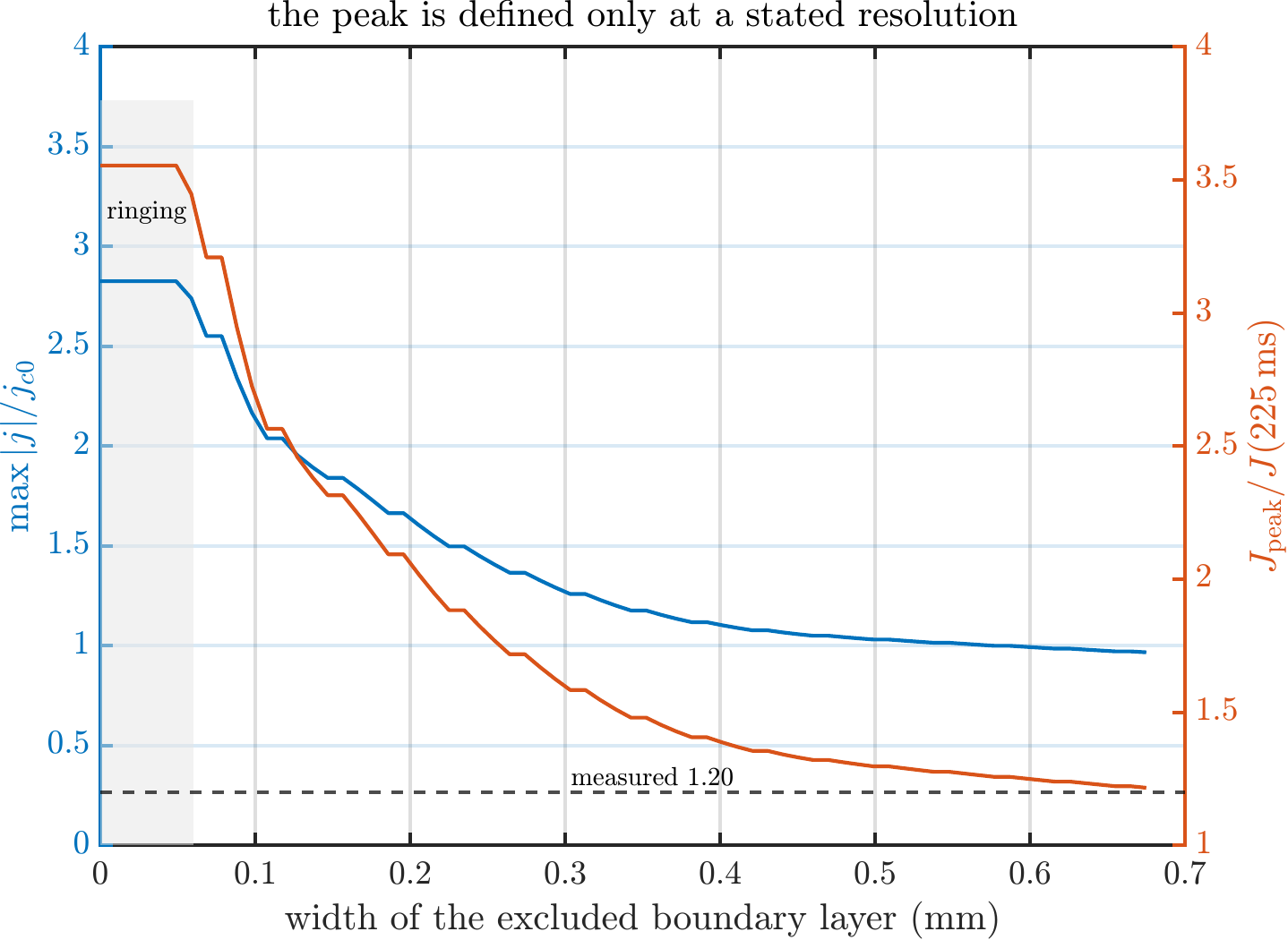}
\caption{The transient peak is defined only at a stated spatial resolution.
Peak current (left axis) and peak-to-asymptote ratio (right axis) as functions
of the width of the boundary layer excluded from the maximum. The shaded strip
is the layer within which the spectral and finite-difference representations
of Faraday's law disagree by more than the signal. The dashed line is the
measured ratio of $1.20$, which is not attained anywhere in the range scanned:
the lowest value is $1.22$, at a guard of $0.68$~mm or $45\%$ of the
half-width. The apparent convergence toward the measured value at large guard
is not agreement, because by then the edge peak has been excluded altogether
and the reported maximum is the central plateau, so the ratio is that of two
plateau values rather than of a peak to its asymptote. For comparison, the
resolution of the magneto-optical indicator is of the order of tens of
microns.}
\label{fig:guard}
\end{figure}

Nor is it instrumental. Figure~\ref{fig:guard} shows the computed ratio
against the width of the excluded boundary layer. It decreases monotonically toward the measured value but does not reach it:
over the whole range scanned, up to $0.68$~mm or $45\%$ of the half-width, the
lowest ratio obtained is $1.22$.

The approach is in any case spurious. Once the excluded layer is that wide the
edge peak has been removed altogether and the reported maximum is simply the
value of the central plateau, so the quantity being compared is the ratio of
two plateau values rather than a peak against its asymptote --- not what the
experiment reports. At every guard for which an edge peak still exists the
ratio exceeds $2$, and the indicator resolution, together with the smoothing
implicit in the Biot--Savart inversion, is of the order of tens of microns,
some twenty times narrower than the guard already used here and three orders
of magnitude below the width at which the curves would meet. Between $0.10$ and $0.15$~mm the curve is
smooth and the ratio is $2.4$--$2.3$, so the conclusion does not depend on
where within that range the guard is placed.

The most obvious escape from this conclusion is that the ceiling in
Eq.~\eqref{eq:law} is itself responsible, and that an unbounded power law
would behave differently. It does not. Raising $\rho_{\max}$ from $2$ to
$10^2$ at fixed $E_c$ does bring the peak-to-asymptote ratio down to $1.17$,
close to the measured value --- but it also moves the turnover to $5.5$~ms,
because $t_{\rm cross}$ depends on the product $\rho_{\max}E_c$ and that
product has been multiplied by fifty. The agreement is an artefact of running
the model four times too fast. Restoring the product by lowering $E_c$ in the
same proportion returns the turnover to $23.7$~ms and the amplitude to
$1.86\,J_{c0}$, within $2\%$ of the bounded-law value. This test was run with
a quite different constitutive parametrisation as well --- creep exponent
$40$, Kim exponent $3$, $B_0=1.33$~T, taken from an earlier implementation of
the same formalism --- so the amplitude is insensitive to that choice too.

The ratio does fall, but the decomposition of
Sec.~\ref{sec:decomp} shows why and it does not help: the steeper field
dependence raises the Kim factor from $1.05$ to $1.37$, and the transient
excess proper is $1.45$ against $1.36$ for the calibrated model. What changes
is how much of the ratio is contributed by $j_c(B)$, not how much transient
current the model produces.

With four observables constraining three principal baseline parameters, the
constitutive test is overdetermined by one observable. The bounded power-law
family fails that independent amplitude test in the direction of excessive
transient current.

\subsection{Decomposing the reported ratio}\label{sec:decomp}
Part of the interpretive difficulty is that the peak-to-asymptote ratio is not
a clean measure of the transient. It factorises as
\begin{equation}
 \frac{J_{\rm peak}}{J_{\rm asym}}
 =\underbrace{\frac{j_c(0)}{j_c(B_{\rm peak})}}_{\text{Kim}}
 \times\ \underbrace{\vphantom{\frac{j_c(0)}{j_c(B)}}\text{creep since the peak}}
 \times\ \underbrace{\vphantom{\frac{j_c(0)}{j_c(B)}}\text{transient overshoot}} ,
\end{equation}
and only the last factor is what the comparison is usually taken to measure.
In the calibrated run the Kim factor is $1.05$. The creep factor has no
limiting value: with a power law there is no true asymptote, $j$ drifts
downward indefinitely, and the ratio grows with however long the observation
lasts. We therefore evaluate it at $225$~ms, matching the experiment, and
recommend that any comparison of this quantity state its observation window.

The decomposition also shows how a large ratio can arise with no overshoot at
all. In a scan of $\rho_{\max}$ at fixed $E_c$, the ratio saturates near
$1.25$ while the amplification falls below unity: the model then reproduces a
$25\%$ excess without any overcritical current. Whether the excess reported
from imaging is transient in origin therefore depends on separating these
contributions, which requires an independently measured $j_c(B)$ and a stated
window.

\subsection{Charge and induction parts of the electric field}\label{sec:helmholtz-results}

\begin{figure*}[!tb]
\centering
\includegraphics[width=\textwidth]{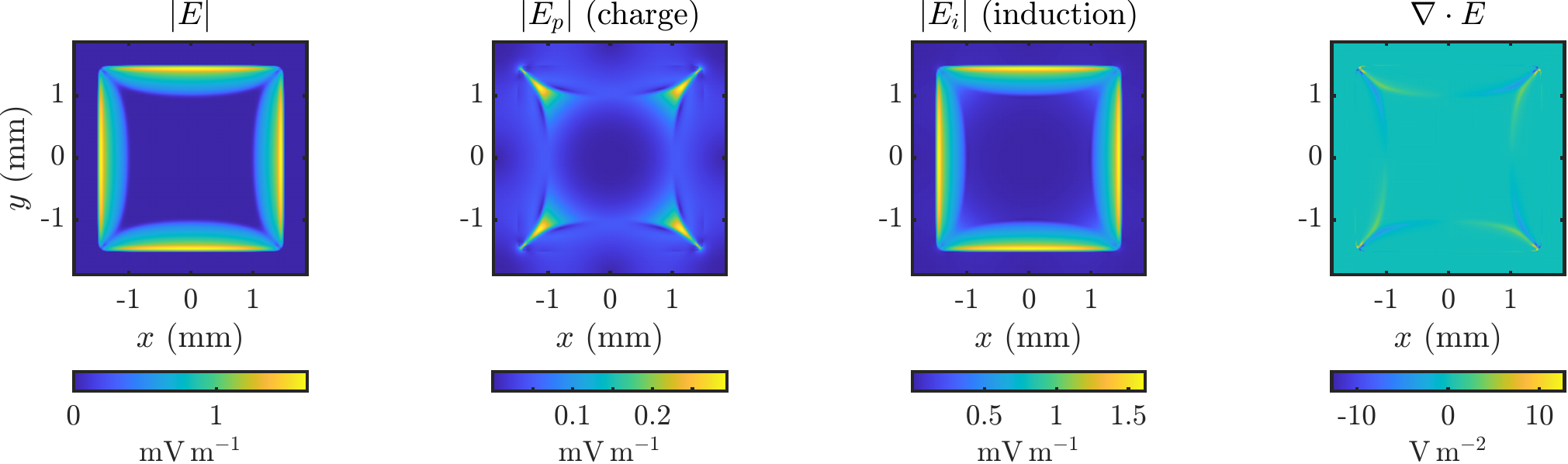}
\caption{Helmholtz decomposition of the computed electric field at the instant
of maximum $|\bm E|$, $18.8$~ms after the start of the ramp. From left to
right: the total field, its irrotational part $\bm E_p=-\nabla\phi$, its
solenoidal part $\bm E_i=\nabla\times(\psi\hat{\bm z})$, and the divergence of
the field. \emph{The colour scales differ between panels}: $|\bm E_p|$ peaks
at $0.28$~mV\,m$^{-1}$ against $1.6$~mV\,m$^{-1}$ for $|\bm E|$, so the
irrotational part would be invisible on a common scale. The solenoidal part
fills the penetrated annulus and reproduces the total field almost exactly,
while the irrotational part is confined to the diagonals and strongest at the
corners. The divergence map identifies these diagonals as the discontinuity
lines: $\nabla\cdot\bm E$ vanishes everywhere else, so the whole of the
transient charge resides there.}
\label{fig:helm}
\end{figure*}

Figure~\ref{fig:helm} shows the four maps at the instant of maximum
$|\bm E|$. The decomposition separates the field into two components with
different geometry and different histories.

The solenoidal part occupies a continuous annulus following the penetrated
region, widest at the centres of the four sides and narrowing toward the
corners, and it reproduces $|\bm E|$ almost exactly: the two maps are
indistinguishable by eye and the temporal maxima of $|\bm E_i|$ and
$|\bm E|$ coincide throughout the simulated interval. This is what
Eq.~\eqref{eq:helm-intro} leads one to expect, since $\psi$ carries the whole
dependence on $\dot B$ and the drive dominates the response.

The irrotational part is a different object altogether. It is concentrated
along the two diagonals, strongest at the corners, forming a cross that is
absent from the total field because it is an order of magnitude weaker. The
divergence map makes the identification unambiguous: $\nabla\cdot\bm E$
vanishes everywhere except on the diagonals, so the whole of the transient
charge sits on the discontinuity lines, where the current direction turns.
This is the configuration inferred indirectly in reconstruction work
\cite{Brandt1995,RomeroSalazar2010}, here obtained from a forward
calculation.

The two parts also peak at different times, and the order is the reverse of
what one might expect. The total field and its solenoidal part reach their
maximum at $18.8$~ms, close to the turnover of the current, whereas
$|\bm E_p|$ peaks at $8.0$~ms --- $10.9$~ms \emph{earlier} --- with a second,
broader maximum near $40$~ms. At the maximum of $|\bm E|$ the ratio is
\begin{equation}
 \frac{\max|\bm E_p|}{\max|\bm E_i|}=0.23 ,
\end{equation}
rising to $0.32$ at its largest over the whole interval, and $|\bm E_p|$ is
negligible beyond about $150$~ms.

A natural reading is that the two components respond to different things. The
solenoidal part tracks the drive and is largest while the flux front is moving
fastest. The irrotational part is largest earlier, while the current pattern
is still reorganising and the discontinuity lines are forming, and it subsides
once the critical-state geometry is established, even though the flux
continues to penetrate. The charge contribution is therefore an early,
subdominant and spatially localised feature, not a residual that outlives the
drive.

We state this explicitly because the opposite is easy to assume. The diagonal
discontinuity lines are the most conspicuous geometrical feature of a square
sample \cite{Schuster1995,Brandt1995}, and the charge that sits on them is the
part of the problem that reconstruction procedures find hardest; it does not
follow that it dominates either the amplitude or the late-time behaviour. In
this calculation it dominates neither, at any time.

\FloatBarrier
\section{Discussion}\label{sec:discussion}
The calculation establishes both the usefulness and the limitation of the
bounded, constant-exponent power-law baseline. Retaining the non-local
thin-film geometry is essential: the model reproduces the observed sequence
of edge screening, turnover, inward displacement, and approach to a Kim-like
profile. The penetration-based sheet current and timing-based dissipation
parameters are mutually consistent, and the calculated relaxation time falls
within the range of the two measured edges. These agreements show that the
overall magnetic geometry and clock can be represented without introducing
sample-specific defects.

The independent amplitude test gives a different result. Once the penetration
and timing constraints are imposed, the calculated peak remains near
$1.9J_{c0}$ throughout the explored power-law parameter space. Changing the
creep exponent moves the relaxation time but not the peak. Moving along the
nearly degenerate $E_c$--$\rho_{\max}$ direction changes the long-time creep
level more than the transient maximum. Varying the Kim field changes how much
of the peak-to-window ratio is attributable to $J_c(B)$, but does not remove
the transient excess. The disagreement is therefore not evidence against the
non-local magnetic formulation; it identifies the constant-exponent
constitutive interpolation as the restrictive part of the baseline.

The measured peak-to-asymptote ratio also requires care. It combines the Kim
field dependence, creep accumulated after the peak, and any genuine transient
overcriticality. Because a power law has no strict long-time asymptote, the
ratio depends on the observation window. A comparison should therefore state
the window and, where possible, use an independently determined $J_c(B)$.
Without that separation, a ratio above unity does not uniquely measure a
transient overshoot.

The edge amplitude is additionally limited by spatial representation. The
ideal sharp mask creates high-wave-number content, and the spectral and
finite-difference forms of Faraday's law disagree within a narrow boundary
layer. Reporting a raw one-cell maximum would therefore assign physical
meaning to a discretization-dependent quantity. Fixing the filter and guard in
physical units makes the guarded amplitude reproducible and exposes the
important conclusion: experimental-scale smoothing cannot reconcile the
power-law amplitude without removing the peak whose amplitude is being
compared.

The Helmholtz decomposition provides a complementary result that is not
available directly from current reconstruction. The solenoidal electric field
follows the penetrated region and dominates the total field, whereas the
irrotational contribution is weaker, localized mainly on the diagonal
discontinuity lines, and reaches its largest value earlier. Thus, the transient
charge structure is geometrically conspicuous but does not dominate the total
dissipation or the late response in this baseline.

Several physical ingredients could move the calculation beyond the present
constitutive family. Local thermal feedback can reduce $J_c$ where dissipation
is strongest. A distribution of activation barriers, or a state-dependent
barrier that evolves with current, induction, and temperature, can decouple
the driven high-mobility stage from the late pinned stage. Spatially varying
pinning may also matter because the imaged specimen contains inhomogeneity and
a visible defect \cite{Wells2017}. These possibilities are motivated by the
baseline discrepancy but are outside the scope of the present study; their
assessment requires a separate formulation and should not be inferred from a
further unconstrained fit of the constant-exponent law.

Ramp-rate dependence offers an experimental route for discriminating among
such mechanisms. Over $\dot B_a=2$ to $64$~T\,s$^{-1}$ at otherwise fixed
baseline parameters, the present model gives
\begin{equation}
 \frac{\max\|\bm J\|}{J_{c0}}\propto\dot B_a^{\,0.22},\quad
 t_{\rm cross}\propto\dot B_a^{-0.66},\quad
 \max\|\bm E\|\propto\dot B_a^{\,0.22}.
 \label{eq:ramp}
\end{equation}
The weak amplitude exponent implies that the experimental uncertainty in the
ramp rate cannot account for the factor-of-two ratio discrepancy. The turnover
also does not scale simply as the inverse ramp rate, showing that it is set
jointly by the drive and the magnetic diffusion time. Measurements at several
ramp rates, together with local thermal information, would therefore test how
the actual vortex mobility departs from the power-law baseline.

Finally, the parameter degeneracy carries a methodological warning. In the
strongly driven regime, $E_c$ and $\rho_{\max}$ enter predominantly through
their product; reporting either quantity without the degenerate combination
can make different parameter sets appear physically distinct when they
produce nearly the same transient. For bounded power-law comparisons, we
recommend reporting $\rho_{\max}E_c$, the creep exponent, the spatial filter,
the boundary guard, the observation window, and the precise peak-tracking
rule.
\FloatBarrier
\section{Conclusions}\label{sec:conclusions}
We have established a non-local, dimensional baseline for transient flux
penetration in a YBCO thin film using a bounded, constant-exponent power-law
constitutive relation. The staged parameter selection reveals that the
turnover time is governed primarily by the product of the criterion field and
the saturation parameter. Those quantities are therefore nearly degenerate
in the driven regime and cannot be identified independently from the turnover
alone; the creep exponent supplies the principal control of the subsequent
relaxation.

With the baseline parameters constrained by the penetration geometry and
timing observables, the model captures the observed three-stage chronology, a
guarded-current turnover near 22~ms, a relaxation constant of 41.2~ms compared
with the measured 37.5 and 43.7~ms, and a late maximum settling approximately
0.92~mm from the edge. The independent amplitude observable is not captured.
At the stated spatial resolution the computed peak is $1.90J_{c0}$ and the
peak-to-225-ms ratio is 2.13, compared with the reported ratio of 1.20. This
excess remains within a few percent under the investigated variations of the
creep exponent, the degenerate dissipation direction, the Kim field
dependence, and the numerical mesh.

The mismatch cannot be removed by experimental-scale smoothing: a boundary
exclusion large enough to approach the measured ratio removes the edge peak
itself and changes the quantity being evaluated. Neither can it be repaired by
changing $E_c$ within a conventional range while preserving the timing. The
conclusion is therefore specific but robust: under the constraints imposed by
the measured geometry and times, the bounded, constant-exponent power-law
family produces excessive transient current.

The analysis additionally shows that the peak-to-window ratio mixes the Kim
field dependence, post-peak creep, and genuine transient overshoot, and must be
reported with a stated observation time. The Helmholtz decomposition shows
that the induction-related electric field dominates the total response,
whereas the charge-related component is weaker and concentrated along the
diagonal discontinuity lines. Finally, the raw edge maximum is not a
resolution-independent observable because the spectral and finite-difference
representations disagree in the immediate boundary layer.

These findings define the role of the present work as a reproducible
power-law reference rather than a complete theory of the experiment. They
identify the constant constitutive exponent and high-current interpolation as
the elements that must be replaced or generalized in models with
state-dependent vortex mobility. The baseline, diagnostics, and failure mode
reported here provide quantitative tests for such extensions.

\section*{Author declarations}

\subsection*{Conflict of interest}
The authors have no conflicts to disclose.

\subsection*{Author contributions}
\textbf{O. A. Hern\'andez-Flores}: Conceptualization (equal); Methodology
(equal); Software (lead); Investigation (lead); Visualization (lead); Writing
-- original draft (lead). \textbf{C. Romero-Salazar}: Conceptualization
(equal); Methodology (equal); Supervision (lead); Writing -- review and
editing (equal). \textbf{F. P\'erez-Rodr\'iguez}: Methodology (supporting);
Validation (lead); Writing -- review and editing (equal).

\section*{Data availability}
The solver, the driver scripts and the processed data that support the
findings of this study are openly available in Zenodo at
\url{https://doi.org/10.5281/zenodo.21882874}, reference number
10.5281/zenodo.21882874. The four MATLAB files reproduce every number and
every figure of this paper from scratch, and the deposit includes a list of
reference values against which a fresh run can be checked.

\section*{Acknowledgments}
The authors acknowledge the institutional support of Universidad Aut\'onoma
Benito Ju\'arez de Oaxaca and Benem\'erita Universidad Aut\'onoma de Puebla.

\appendix
\section{FFT-compatible grid including the sample boundary}\label{app:grid}
With $x_i=(i-1)\Delta x-a$ and $\Delta x=2a/N_x$, requiring a node to
coincide with $x=\pm1$ gives $k=1+\tfrac{N_x}{2}(a\pm1)/a$. Writing $a=p/q$
with positive integers, the choice $N_x=2pN$ gives integer boundary indices
$k_\pm=1+N(p\pm q)$. The integers are also chosen so that $N_x$ factorises
efficiently for the FFT.

\section{Fourier multipliers and the gauge}\label{app:gauge}
The multiplier $2/\lVert\bm k\rVert$ in Eq.~\eqref{eq:ginv} and
$1/\lVert\bm k\rVert^2$ in Eq.~\eqref{eq:helm-fourier} are set to zero at
$\bm k=0$, and the residual constant is fixed by
$\int_{\Omega_{\rm out}}g\,\mathrm d^2r=0$. One consequence is worth recording
because it is a trap: at $t=0$ there is no current, $\dot H_z$ vanishes inside
the film, and with a zero exterior guess $\dot H_z-\dot H_a$ is a constant
whose transform lives entirely at $\bm k=0$. Annihilating that mode gives
$\dot g\equiv0$ --- a spurious fixed point at which the film never begins to
screen. What breaks it is the mean-value constraint of
Ref.~\cite{Prigozhin2018}, which must therefore be applied to the initial
guess and not only inside the iteration.

\FloatBarrier
\bibliographystyle{aipnum4-2}
\bibliography{Transient_bibfile}
\end{document}